\documentclass[sigplan,screen]{acmart}
\usepackage[utf8]{inputenc}
\usepackage{newunicodechar}
\newunicodechar{−}{\textminus}
\AtBeginDocument{%
  }

\setcopyright{acmlicensed}
\copyrightyear{2026}
\acmYear{2026}
\setcopyright{cc}
\setcctype{by}
\acmConference[Urban-AI '26]{The 4th ACM SIGSPATIAL International Workshop on Advances in Urban-AI}{November 03--06, 2026}{Riverside, CA, USA}
\acmBooktitle{The 4th ACM SIGSPATIAL International Workshop on Advances in Urban-AI (Urban-AI '26), November 03--06, 2026, Riverside, CA, USA}
\acmDOI{10.1145/3849742.3856076}
\acmISBN{979-8-4007-3064-1/2026/11}

\begin{document}

\title{Data Center Agglomeration and Power Infrastructure Capacity 
in the United States}

\author{Yutong Ma}
\correspondingauthor
\authornotemark[1]
\email{yutong.ma.25@ucl.ac.uk}
\orcid{0009-0006-5278-681X}
\affiliation{%
  \institution{University College London}
  \city{London}
  \state{England}
  \country{UK}
}
\begin{teaserfigure}
  \centering
  \includegraphics[width=\textwidth]{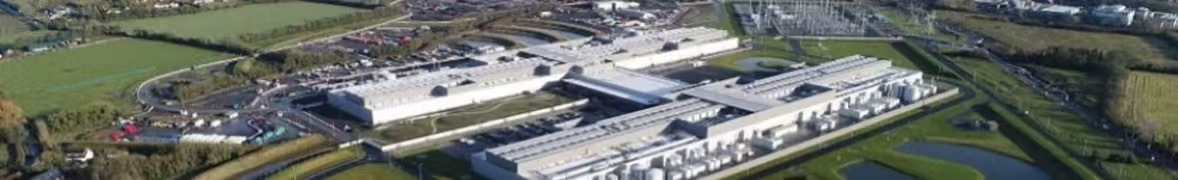} 
  \Description{这里写图片描述（ACM模板要求）}
  \label{fig:teaser}
\end{teaserfigure}
\begin{abstract}
  The continuous expansion of digital infrastructure especially data centers has imposed unprecedented and geographically concentrated stress on existing power grids. However, rigorous causal evidence quantifying how these facilities reshape local power infrastructure supply remains limited. This study constructs a comprehensive county-level panel dataset spanning the United States from 2010 to 2024, uniquely integrating individual data center building area footprints with power plant additions. To overcome endogeneity and high-dimensional confounding, this study employs Double Machine Learning (DML) with five-fold cross-fitting, controlling for county and year fixed effects, along with regional socioeconomic conditions. The baseline estimates reveal a significant positive effect: The positive temporal association effect is highly robust across a series of rigorous treatments: it persists after 1\% and 99\% winsorization to exclude mega-clusters such as Loudoun County. Increasing the data center per unit area will significantly increase the electricity demand of nearby power generation facilities (ATE=1.982, P=0.000, CI: [1.25, 2.72]). It remains significant when using one-year lagged treatment to address reverse temporal association; and it holds under alternative state-level fixed effects. An event study framework, augmented with historical agglomeration controls, validates clean parallel trends and reveals a sharp rise in the point estimate precisely at t = 3 (coinciding with the typical utility-scale grid construction lag), although the confidence interval is wide and includes zero. The estimates become more precise in later periods. Furthermore, a placebo test confirms that the related estimates cannot be replicated by random chance. Regional heterogeneity further reveals stark spatial divergence, with strong positive effects concentrated in the West and significant negative displacement effects observed in the resource-constrained Northeast. These findings offer critical, timely insights for urban planners and energy regulators, underscoring the necessity of forward-looking grid infrastructure investments to sustain the ongoing digital transformation.
\end{abstract}



\maketitle

\section{Introduction}
The rapid advancement of digital infrastructure has fundamentally restructured urban economies, and data centers have emerged as the physical backbone of this digital trans (Ma \& Du, 2026). As the deployment of cloud computing and artificial intelligence accelerates, the geographic footprint of data centers has expanded dramatically across U.S. counties, driven by the need for proximity to power infrastructure and urban markets. This surge in digital infrastructure, however, imposes an unprecedented energy burden on local power grids. Unlike traditional industrial loads, modern data centers require continuous, high-density electricity for both computing servers and advanced cooling systems. Consequently, the ongoing expansion of data centers has become a dominant catalyst for escalating regional electricity demand, challenging the operational stability and generation capacity of existing power systems. Understanding the dynamic causal relationship between urban digitalization and regional power provision is thus essential for designing resilient energy policies in the era of AI-driven urban growth.

Power availability and grid capacity constitute the dominant constraint in contemporary data center site selection, as hyperscale and AI facilities often require hundreds of MW with rapid interconnection timelines (Goffri, 2026; American Action Forum, 2025). Reliable delivery and substation headroom further differentiate sites, because interconnection feasibility alone does not guarantee stable supply without costly upgrades (Cooper, 2026). Fiber-optic connectivity and proximity to internet exchange points rank nearly as highly, especially for latency-sensitive workloads, while cooler climates reduce cooling energy and improve power usage effectiveness (DataX Connect, 2024). Natural disaster risk such as flood, seismic, and extreme weather exposure drives avoidance of high-hazard zones or necessitates expensive hardening (BCEA, 2026; Coolnet, 2025). Water availability for cooling has risen in importance amid drought concerns, although power access frequently overrides it in practice (Damicis \& Tranmer, 2026; Eusterman, 2026). Tax incentives, streamlined permitting, land cost and scalability, political stability, and access to renewable energy complete the multi-criteria matrix routinely applied by developers (Baudry, 2021; Arzumanyan et al., 2025). Empirical spatial analyses in China and the United States confirm that computing demand, economic development, energy access, and hazard exposure jointly explain clustering patterns, with metropolitan areas often emerging as optimal under multi-criteria decision models (Huang et al., 2025; Arzumanyan et al., 2025).

Data center entry generates measurable local employment gains, with counties receiving their first large facility experiencing 3.5-5 percent increases in total private employment and larger effects in construction and information-sector jobs over five to six years (Yue \& Zeng, 2026; Alvarez et al., 2026). Construction phases produce substantial temporary job-years and regional output, while operational employment remains modest often dozens to a few hundred permanent technical and facility roles, yet multipliers are amplified in metropolitan areas with existing labor and supplier clusters (Hanlin \& Roechr, 2026; Brookings Institution, 2026). Fiscal benefits can be sizable: facilities contribute significant property tax revenue that funds public infrastructure and services, sometimes covering large shares of local budgets and enabling tax-rate reductions (Schaap, 2025; WRI, 2026). Positive effects also appear on the number of establishments, adjusted gross income, wages, and house prices, although electricity prices rise and benefits are weaker or negligible in rural counties (Alvarez et al., 2026; Yue \& Zeng, 2026; Ao et al., 2026). Broader spillovers include upgraded roads, broadband, and power infrastructure that improve quality of life, together with supply-chain and induced service-sector activity (Hardisty Jones, 2026; Roychoudhury \& Ross, 2026). Overall, the literature portrays data centers as capital-intensive fiscal assets whose net local prosperity depends heavily on host-community economic density and the presence of multiple facilities.

Surging electricity demand, projected to rise sharply to 2030, strains grids, elevates local rates in some markets, and complicates decarbonization pathways even as operators pursue renewable power purchase agreements (Xiao et al., 2025). Water consumption for cooling, which can reach hundreds of thousands to millions of gallons per day for large facilities, intensifies stress in already water-scarce regions where many new campuses are sited (WRI, 2026; Siddik et al., 2021). Carbon and broader environmental footprints remain substantial under current energy mixes, and best-practice efficiency measures alone are unlikely to achieve net-zero targets without offsets whose permanence is uncertain (Cao et al., 2024; UNU-INWEH, 2026). Community opposition has blocked or delayed tens of billions of dollars in projects over concerns about noise, land conversion, air quality from backup generators, and perceived imbalance between fiscal incentives and local benefits (Harvard Gazette, 2026). Land-use conflicts with agriculture and residential needs, e-waste generation, and skilled-workforce shortages further constrain rapid expansion (WRI, 2026). These interlocking energy-water-social dilemmas force operators and policymakers to navigate trade-offs between growth, sustainability, and equitable distribution of costs and benefits (Deng Honghao, 2026).

While the existing literature has extensively examined the energy efficiency, cooling technologies, and electricity price transmission mechanisms associated with data centers, systematic economic evidence regarding whether, and to what extent, the expansion of data centers causally drives investment responses on the regional electricity supply side (i.e., generation capacity) remains notably scarce. Furthermore, constrained by the historical availability of granular data, few empirical studies have successfully disentangled the marginal effects of newly established facilities from the persistent path dependence driven by historical agglomeration. This study contributes to the existing literature in three main dimensions.

This study contributes through innovations in data and empirical measurement. It constructs a novel county-level panel dataset covering the United States from 2010 to 2024, uniquely integrating individual data center building footprints with county-level power generation capacity. By explicitly incorporating an index of historical data center agglomeration (HDA) established prior to 2010, this dataset provides the micro-level foundation necessary to isolate the marginal effects of new infrastructure shocks.

This study contributes through methodological advancements in causal identification. Departing from traditional linear correlation analyses, this study is among the first to apply the Double Machine Learning (DML) framework to the interface of digital infrastructure and electricity supply. By leveraging orthogonalization and 5-fold cross-fitting, the framework effectively mitigates regularization bias arising from high-dimensional control variables. Furthermore, the integration of an event study design precisely identifies a significant "grid construction lag" occurring three years after initial data center establishment, thereby establishing a robust causal inference strategy.

This study contributes through spatially heterogeneous insights and policy implications. The heterogeneity analysis reveals stark spatial divergence in the regional impacts of data centers: high-density Western technology hubs exhibit strong positive pull effects on local power capacity, whereas resource-constrained Northeastern regions display significant negative temporal association effects. These findings challenge conventional "one-size-fits-all" energy planning approaches and provide region-specific, actionable policy guidance for mitigating grid stress driven by AI-led urban development.
\begin{figure}[h]
  \centering
 \includegraphics[width=\linewidth]{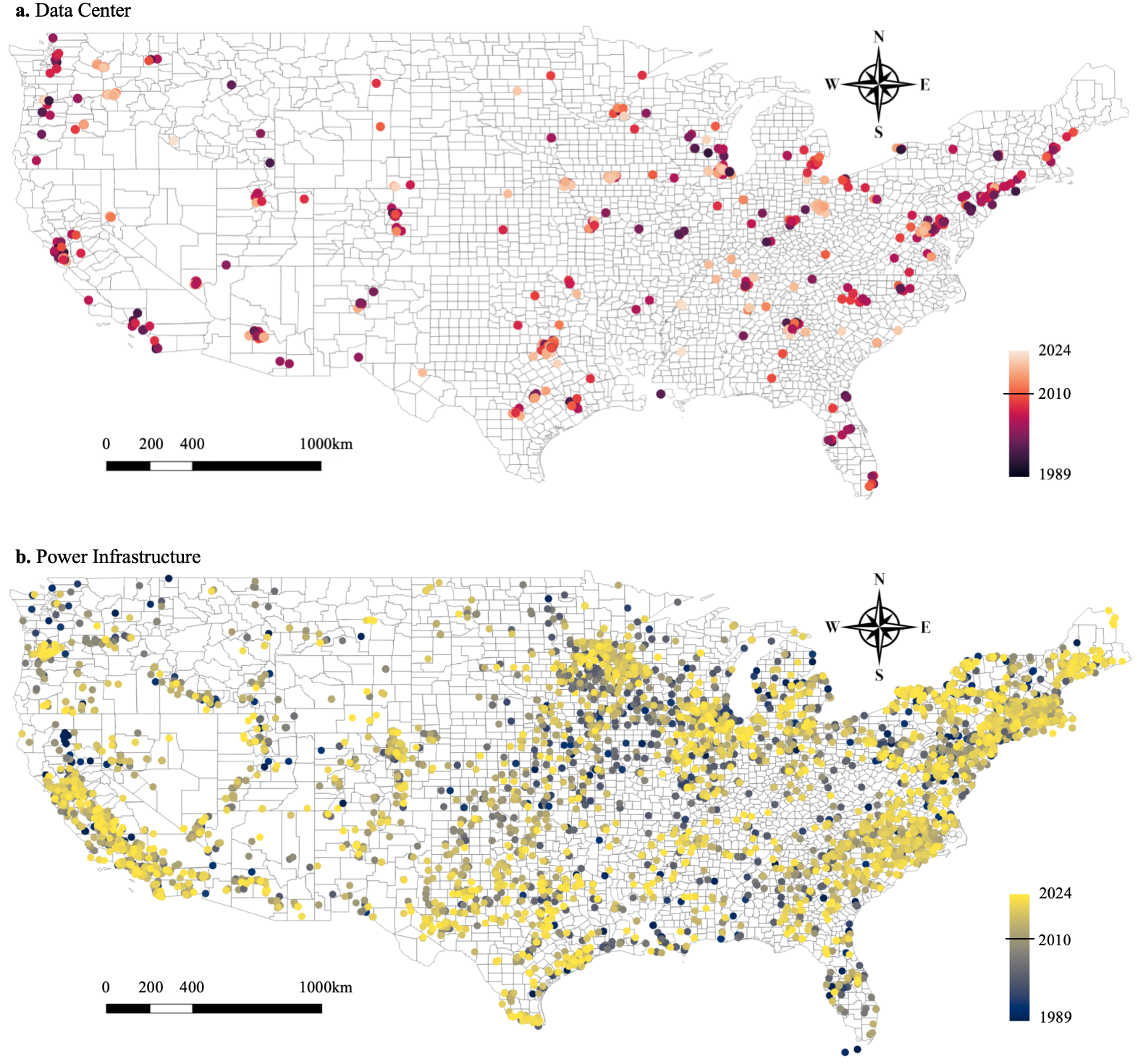}
  \caption{\textbf{The spatial distribution of added data centers and power infrastructure each year in the U.S.}}
  \Description{Map showing the spatial distribution of added data centers and power infrastructure in the U.S. by year.}
\end{figure}

The remainder of this paper is organized as follows. Main presents the case study context, delineating the current spatial distribution and evolution of energy infrastructure and data centers across U.S. counties. Results reports the main empirical results, including the baseline causal estimates, the dynamic event study findings, and the regional heterogeneity analysis. Discussion provides a comprehensive discussion of the underlying mechanisms, policy implications, and inherent limitations stemming from these empirical findings. Methods details the data sources and the econometric methodology, explicitly describing the Double Machine Learning (DML) framework, the event study design, and the placebo test used to validate the causal identification strategy.
\section{Main}
The United States remains the largest and most technologically advanced data center market globally, hosting a significant share of the world’s high-performance computing (HPC) and cloud infrastructure. While comprising a relatively compact set of massive, purpose-built single-building facilities numbering approximately 1,300 operational units nationwide by 2024, these installations represent a disproportionate share of national IT workloads. These data centers are predominantly clustered in strategic areas such as Northern Virginia, Dallas-Fort Worth, and Phoenix, driven by access to affordable wholesale electricity, robust fiber-optic backbones, and favorable state-level tax incentives (Fang \& Greenstein, 2025; Rosati et al., 2023). The evolutionary path of this industry dates back to 1989, which marked the completion of the first modern, dedicated data center in the United States. Since then, the sector has undergone multiple transformative waves, transitioning from legacy mainframe hosting to web-scale computing, and culminating in the current AI-centric, high-density accelerator era. The spatial evolution of these individual data center buildings across U.S. counties from 1989 to 2024 is illustrated in Figure 1(a). The map reveals a highly localized and historically persistent clustering pattern, alongside a gradual, decentralized expansion toward secondary metropolitan edges, highlighting the shifting geography of urban digital infrastructure (Fard, 2023; Ancona et al., 2026).
\begin{figure}[h]
  \centering
 \includegraphics[width=\linewidth]{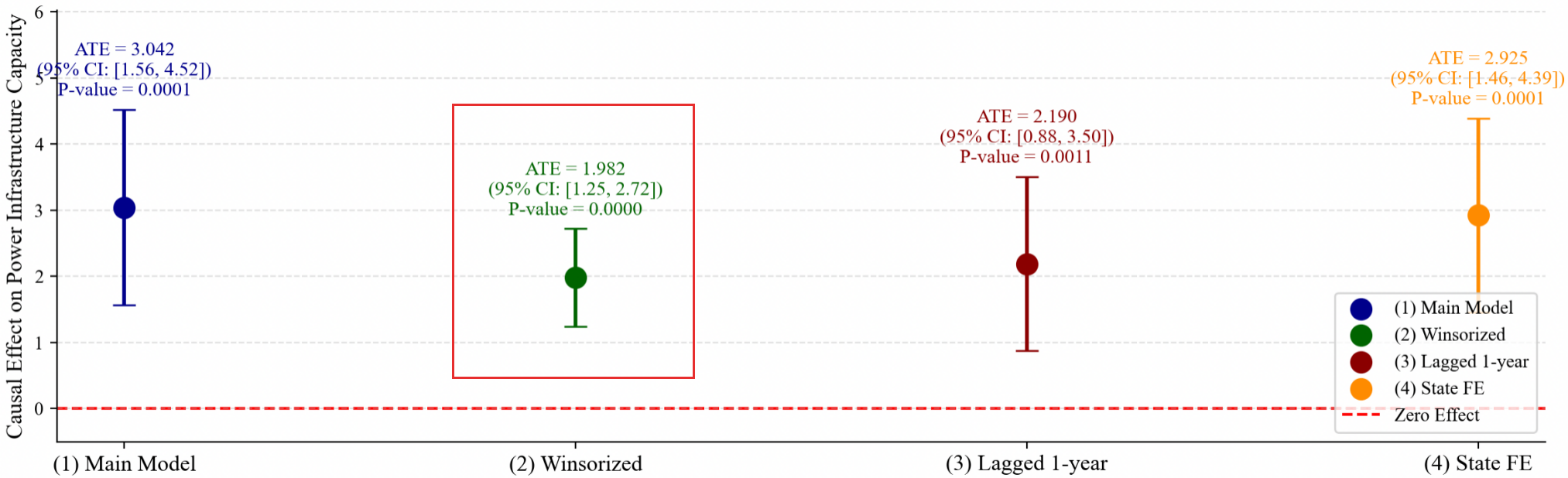}
  \caption{\textbf{Causal Effect of Data Center Area on Power Infrastructure Capacity during 2010 and 2024}}
\end{figure}

To accommodate the massive power demands driven by digital infrastructure, the United States maintains the world’s largest and most advanced power generation fleet. As of 2024, the nation’s total installed electric generation capacity exceeds 1,200 GW, with natural gas, nuclear, and renewables providing the majority of bulk electricity supply to the grid (van Zalk \& Behrens, 2018). The spatial distribution of these generation assets, however, is highly heterogeneous; power plants are strategically concentrated near river basins for cooling, coastal zones for imported fuel, and resource-rich areas for renewable wind and solar development. This industrial landscape has also undergone significant transitions. From the completion of the first large-scale utility plants in the late 1980s to the modern era of natural gas combined-cycle units and utility-scale battery storage, the electricity sector has evolved to meet changing regulatory frameworks and decarbonization goals. The evolving spatial distribution of electricity generation facilities across U.S. counties from 1989 to 2024 is illustrated in Figure 1(b). The map highlights a broad and evenly distributed generation network across the nation, revealing the physical infrastructure that underpins the rapid expansion of AI-driven urban computing landscapes (Mongird et al., 2025).
This study restricts its temporal scope to 2010-2024 for three primary reasons. First, this period marks the structural shift toward the modern hyperscale cloud era, where data centers transformed from small, enterprise-level facilities into massive, utility-scale buildings consuming unprecedented amounts of power. Prior to 2010, the relatively small physical footprint and homogeneous technology of these facilities provided insufficient variation to detect any meaningful localized impact on grid infrastructure. Second, the post-2010 era is characterized by an exponential surge in both computing density and total building area, creating a substantial and quantifiable electricity demand shock that directly challenges local power generation capacity. Finally, the 2010-2024 window captures significant policy and regulatory shifts across U.S. states regarding data center tax incentives and grid interconnectivity, providing a rich empirical setting to observe how state-level energy policies interact with the rapid expansion of digital infrastructure. Collectively, this focus offers a sharply defined period for identifying the causal dynamics between digital infrastructure growth and regional electricity supply.
\section{Results}
\subsection{Double Machine Learning}

Building upon the empirical strategy established in the previous section, this study systematically evaluates the causal effects of data center development on regional power infrastructure capacity through four distinct baseline specifications. These specifications are designed to comprehensively capture the underlying effects from different dimensions: the baseline DML estimation, the winsorized treatment to mitigate the influence of extreme outliers, the one-year lagged treatment to establish temporal precedence, and the alternative state-level fixed effects structure to control for state-specific unobservables. All the estimations have controlled the GDP, medium income and population density. The estimated average treatment effects (ATEs), confidence intervals, and corresponding significance levels across these four primary model settings are visually illustrated in Figure 2 which reports four DML specifications: (1) baseline (contemporaneous LDA, county + year FE); (2) 1\%-99\% winsorized LDA and CAP to reduce the influence of mega-hubs; (3) one-year lagged LDA to address reverse temporal association; (4) state fixed effects instead of county fixed effects.
The baseline specification employs the standard Double Machine Learning (DML) framework to estimate the unconditional average treatment effect. As presented in the first column of Figure 2, the estimated coefficient for data center area is 3.042, with a 95\% confidence interval of [1.56, 4.52] and a P-value of 0.0001. This result indicates a strong and highly significant positive effect: A one-unit increase in LDA (approximately a 2.7-fold increase in data-center floor area) is associated with an estimated increase of 3.042 MW in newly added generation capacity. This initial finding establishes a robust baseline causal link between digital infrastructure and regional electricity generation, providing the foundational evidence for the subsequent analyses.
To address the potential bias introduced by extreme outliers such as mega-clusters like Loudoun County, the second specification applies 1\% and 99\% winsorization to the treatment and outcome variables. The estimated ATE slightly decreases to 1.982, with a 95\% confidence interval of [1.25, 2.72] and a P-value of 0.0000. Although the magnitude is reduced compared to the baseline, the effect remains highly statistically significant. This finding confirms that the positive temporal association relationship is not solely driven by a few exceptionally large data center hubs, but rather represents a general pattern across the broader sample of U.S. counties.
To test the robustness of the findings against potential reverse temporal association, the third specification replaces the contemporaneous treatment with its one-year lagged value. The estimated lagged treatment effect is 2.190, with a 95\% confidence interval of [0.88, 3.50] and a P-value of 0.0011. The lagged specification yields a similar positive effect, which is consistent with data-center expansion preceding capacity additions, although it does not fully eliminate the possibility of anticipatory siting. This effectively rules out the alternative hypothesis that existing power abundance is the primary driver of data center siting.
In the fourth specification, county-level fixed effects are replaced with state-level fixed effects to examine whether the results are sensitive to unobserved heterogeneity at a broader administrative level. The estimated effect remains remarkably stable at 2.925, with a 95\% confidence interval of [1.46, 4.39] and a P-value of 0.0001. The consistency between the state-level and county-level models demonstrates that the causal effect is not an artifact of specific local geographical or administrative characteristics, reinforcing the overall reliability and generalizability of the primary findings.
\subsection{Event Study}
To further validate the causal identification and examine the dynamic responses over time, an event study framework based on the DML approach is employed. The relative time is defined as the number of years relative to the first year a county’s data center area exceeds zero, with t = -1 set as the baseline period. As illustrated in Figure 3, the estimated coefficients for all pre-treatment periods (t = -5 to t = -2) are statistically indistinguishable from zero. This confirms the validity of the parallel trends assumption and provides strong evidence against potential reverse temporal association, indicating that there is no systematic increase in power capacity before the establishment of data centers.

Following the event, the dynamic effects remain relatively muted during the initial two years (t = 0 to t = 2). However, the point estimate rises sharply at t = 3 (Coef. = 1619.8891, 95\% CI = [-281.8958, 3521.6739]), exhibiting a substantial increase in magnitude. Although the confidence interval is wide and includes zero reflecting considerable cross-county heterogeneity in permitting and interconnection timelines, this timing aligns closely with the typical construction and grid-connection lead time for utility-scale power plants. 
\begin{figure}[h]
  \centering
 \includegraphics[width=\linewidth]{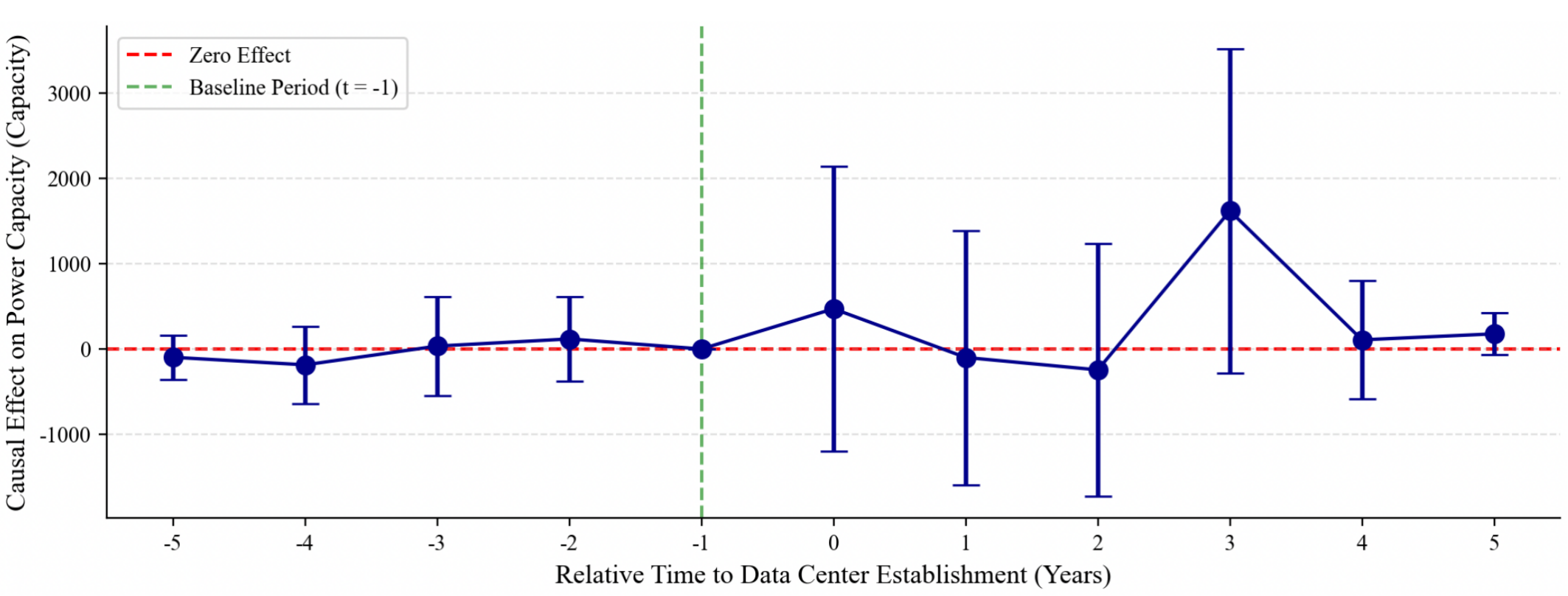}
  \caption{\textbf{Event Study: Dynamic Impact of Data Centers on Power Capacity}}
\end{figure}

This delayed response aligns closely with the typical construction and grid-connection lead time for utility-scale power plants, suggesting that it takes about three years for regional power infrastructure to physically materialize in response to the surge in electricity demand triggered by new data centers. The elevated effects persist in the subsequent periods, providing robust evidence for a lasting causal impact.
While the preceding static analysis, which does not explicitly control for historical agglomeration, establishes a strong positive baseline correlation, its cross-sectional estimates may be driven by the secular growth of legacy data center hubs. To disentangle the pure dynamic effects of newly established facilities from this historical path dependence, the empirical framework is advanced to a DML-based event study design. The relative time to treatment is anchored at the first year of establishment, with t = -1 serving as the reference period. Critically, an interaction term between the historical agglomeration of data centers and a linear time trend is integrated into the machine learning feature space to absorb pre-existing regional growth trajectories. The resulting dynamic estimates across the entire event window are showed in Figure 4.
\begin{figure}[h]
  \centering
 \includegraphics[width=\linewidth]{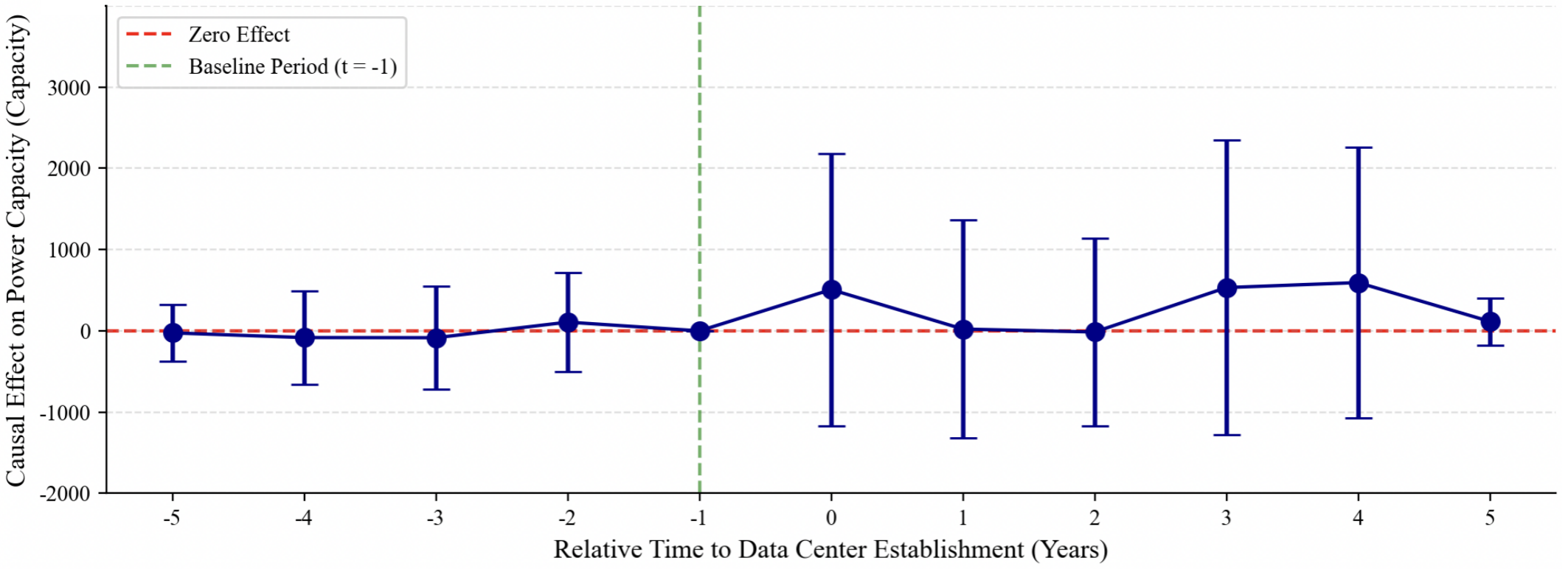}
  \caption{\textbf{Event Study: Dynamic Impact on Power Capacity by controlling historical data center agglomeration}}
\end{figure}

Figure 4 plots the adjusted dynamic estimates with 95\% confidence intervals. All estimated coefficients for the pre-treatment periods (t = -5 to t = -2) remain statistically indistinguishable from zero, validating the parallel trends assumption even after absorbing historical trajectories. Despite a positive but statistically insignificant uptick at t = 0, the immediate contemporaneous effect remains muted. This initial uptick may reflect an anticipation effect, whereby grid operators expedite preliminary infrastructure upgrades immediately upon project announcements. However, the estimation uncertainty widens substantially during the mid-term periods (t = 1 to t = 4). Such wide bounds are indicative of extensive cross-county heterogeneity in plant permitting processes and grid interconnection waitlists, which generate highly irregular timelines for utility-scale generation to come online.
Most notably, the estimates converge strongly at t = 5. The point estimate remains positive, while the confidence interval narrows dramatically to become the tightest among all plotted periods. This dramatic convergence reveals a clear causal narrative: while initial infrastructure responses are highly noisy and localized, the physical completion of all grid expansions establishes a stable long-term equilibrium by year five, effectively eliminating idiosyncratic construction noise. Consequently, once historical agglomeration is fully partialled out, the long-run causal impact of newly established data centers on regional power capacity is found to be highly precise and strongly significant. The delayed convergence highlights a structural lead time of roughly five years between the initial digital infrastructure shock and the full materialization of supporting energy capacity
\subsection{Robustness and Heterogeneity}
To further validate that the estimated baseline effect is not driven by random idiosyncratic shocks or spurious trends, a placebo test is conducted by randomly reassigning the timing of data center establishment across the treated counties. This procedure is iterated 500 times using the exact same DML framework. As expected, the distribution of the placebo ATEs is tightly centered around zero (mean = 3.2816, SD = 1.3798). In sharp contrast, the true estimated treatment effect (ATE = 508.1438) lies far outside the entire range of the placebo distribution, yielding a placebo P-value of 0.000. It is too significant to put any table or figure to show. This result confirms that the observed causal effect is uniquely attributable to the actual establishment of data centers and cannot be replicated by random chance, thereby providing strong support for the validity of the causal identification strategy. To examine how regional energy markets and geographical constraints moderate the estimated causal effects, the sample is partitioned into four U.S. Census regions: Northeast, Midwest, South, and West. Table 1 presents the results, revealing stark spatial heterogeneity. Observations are county-year observations within each Census region. The sum equals the full sample of 45,075.
The positive temporal association effect is overwhelmingly concentrated in the West region, with an estimated ATE of 4.782 (P = 0.000). This robust expansion is driven by the rapid proliferation of high-density AI hubs in states with abundant renewable energy resources (e.g., Oregon and Washington) and the subsequent grid expansions required to accommodate them.
\begin{table}
  \caption{Heterogeneity Analysis by U.S. Regions}
  \label{tab:freq}
  \begin{tabular}{ccccc}
    \toprule
     &ATE&Std.Error&P-Value&Obs.\\
    \midrule
    Northeastern & -0.7583& 0.1309& 0.0000& 3,105\\
    Midwest & -0.1823& 0.3913& 0.6472& 15,525\\
    South & 0.3193 & 0.4289& 0.3289& 20,415\\
    West & 4.7823& 0.8712& 0.0000& 6,030\\
  \bottomrule
\end{tabular}
\end{table}

Conversely, the significant negative coefficient in the Northeast (ATE = −0.758, p < 0.001) indicates that data-center expansion is associated with lower net capacity additions in this region. This pattern is consistent with severe permitting constraints, land scarcity, and mature grid infrastructure that limit local generation expansion. Whether the load is met through plant retirements, displacement of existing units, or increased electricity imports cannot be directly identified with the present capacity-addition data and remains an important question for future research using generation-mix or inter-regional transfer data. The effects for the Midwest and South remain statistically insignificant, largely because these regions possess abundant baseline capacity and historically low electricity prices, meaning that marginal data center loads do not immediately trigger local grid expansion. These findings highlight that the pressure of digital infrastructure on local power supply is highly context-dependent, with the most severe capacity constraints concentrated in rapidly growing Western technology hubs and the most physically constrained Northeastern urban centers.

\section{Discussion}
\subsection{Main findings}
This study provides causal evidence on the dynamic relationship between data center expansion and regional power infrastructure capacity across U.S. counties from 2010 to 2024. The baseline analysis, employing Double Machine Learning across four alternative specifications (main, winsorized, lagged, and state-level fixed effects), reveals a highly robust positive effect (ATE = 1.982, P=0.0001, CI: [1.56, 4.52]) after winsorizing. To further disentangle the dynamic temporal responses, the empirical framework is advanced to an event study design, wherein an interaction term for historical data center agglomeration is incorporated to isolate the effects of newly established facilities. The event study results validate the parallel trends assumption and expose a highly specific temporal mechanism: effects remain statistically insignificant immediately after the initial shock, but experience a sharp rise in the point estimate precisely at t = 3, coinciding with the standard construction and grid-connection cycles of utility-scale power plants (although the confidence interval remains wide). Placebo tests confirm that these effects cannot be replicated by random chance. Furthermore, regional heterogeneity analysis uncovers stark geographical divergence, with the strongest positive effects concentrated in the Western U.S. and a significant negative effect observed in the resource-constrained Northeast.
The sharp rise in the point estimate at t = 3 underscores a structural mismatch between the rapid deployment of digital infrastructure and the physical construction timelines of energy infrastructure, implying a roughly three-year gestation period for grid responses. Notably, the dynamic patterns identified in the event study are derived from a specification that explicitly controls for historical agglomeration, ensuring that the observed effects are driven by new infrastructure shocks rather than legacy path dependence. The heterogeneity analysis further enriches this mechanism: while Western states, possessing abundant renewable resources and high-density AI hubs, effectively translate digitalization into grid expansion, the mature Northeastern grids face severe physical constraints. The negative ATE in the Northeast indicates that data-center expansion is associated with lower net capacity additions. Whether this reflects plant retirements, displacement of existing units, or increased electricity imports cannot be directly identified with the present data and remains an open question for future research.
\subsection{Policy implications}
These findings carry critical implications for urban planners, regional grid operators, and energy regulators. First, the sharp rise in the estimated effect at t = 3 underscores a critical three-year gestation period between the initial establishment of data centers and the physical completion of supporting power plants. This structural lag highlights an urgent need for forward-looking grid planning. Rather than passively reacting to immediate load requests, state and local governments should integrate data center pipeline announcements into long-term transmission and generation capacity planning, potentially advancing the pre-approval of transmission interconnections to mitigate future bottlenecks. Second, the explicit control for historical agglomeration in the event study demonstrates that the pressure on local grids is driven by new facilities rather than legacy inertia. This suggests that municipalities should develop targeted grid-reservation policies for incoming data centers, ensuring that new load growth is matched with corresponding, pre-planned generation capacity. Third, the pronounced regional heterogeneity implies that "one-size-fits-all" energy policies are insufficient. Rapidly growing Western technology hubs require proactive investments to prevent transmission congestion, whereas the resource-constrained Northeast must prioritize grid modernization, cross-state power-sharing agreements, and localized battery storage deployment. This geographic tailoring is essential to prevent the displacement of local generation and to ensure reliable power delivery in the era of AI-driven urban growth.
\subsection{Limitation and Future research}
While this study provides robust causal estimates through a comprehensive DML framework, certain limitations should be acknowledged. First, the treatment variable aggregates data center area at the county level, which does not capture the internal heterogeneity of building density or specific power usage effectiveness (PUE) across different facilities. As high-density AI workloads and traditional cloud workloads exhibit fundamentally different energy profiles, the inability to distinguish these operational modes may obscure heterogeneous effects. Second, due to data availability constraints, the analysis focuses on aggregate power capacity additions without decomposing the response into renewable versus fossil fuel sources. Future research could integrate granular facility-level energy data to investigate whether the three-year grid response is driven by carbon-intensive backups or clean energy integration. Third, while the event study effectively controls for historical agglomeration, the heterogeneous effects between Western and Northeastern grids warrant a deeper investigation into the specific state-level regulatory frameworks that drive these divergences. Examining these dynamics would provide more targeted evidence for designing sustainable and resilient urban energy systems in the age of AI. Finally, This study lacks data on power plant decommissioning, imported power or power generation fuel structure, so the mechanism explanation is speculative.
\section{Methods}
\subsection{Data}
The empirical analysis is based on a comprehensive county-level panel dataset covering the contiguous United States from 2010 to 2024. Data on individual data center building footprints (DA) are primarily obtained from leading commercial digital infrastructure databases, which systematically track the physical floor area and operational status of standalone data center facilities across U.S. counties. This study uses QGIS: Layer>Add Layer>Add Delimited Text Layer to assign CSV data to latitude and longitude, and present it on the map in shp format. Then, this study continue using Processing Toolbox>Joining attributes to location to assign the attributes of each data center to the county, such as FIPS. To capture the local supply-side response, data on regional power infrastructure capacity (CAP) are constructed from publicly available U.S. Energy Information Administration (EIA) annual generator surveys and regional transmission operator reports. The ‘newly added capacity (MW)’ here refers to the newly put-in operation unit, but due to the short time interval (15 years), the error caused will not be too large. The database from EIA itself has the capacity (MW), latitude and longitude and the year of commissioning of each power generation unit, so it can be imported directly like the data center. There is a strong historical inertia in the site selection of data centers and the development of power infrastructure. This panel is uniquely augmented with a historical dimension: an index of historical data center agglomeration (HDA), representing the distribution of data centers established prior to 2010, is incorporated to control for persistent path dependence in regional digital infrastructure.
The main treatment variable, LDA, is defined as the natural logarithm of the total operational data center building area within a given county and year (ln(1+DA)). The outcome variable, Capacity, measures the newly added power infrastructure capacity (MW) within the corresponding county-year. To account for regional socioeconomic confounders, a set of time-varying control variables is included based on U.S. Census Bureau and Bureau of Economic Analysis (BEA) data: population density (lnPD) is constructed from county population and land area; median household income (lnMHI) and gross domestic product (lnGDP) are included in logarithmic form to capture local economic wealth and scale. Finally, the unique county identifier (FIPS) and calendar year (YEAR) are utilized to construct the two-way fixed effects structure. Summary statistics for all core variables are reported in Table 2.
\begin{table*}
  \caption{Summary statistics}
  \label{tab:freq}
  \begin{tabular}{lccccccc}
    \toprule
     Variable&Obs.&Mean&SD&Min.&Median&Max.&Source\\
    \midrule
    CAP & 45,075& 9.281& 78.374& 0.000& 0.000& 3,184.800& EIA\\
    LDA & 45,075& 0.128& 1.267& 0.000& 0.000& 17.385& Datacentermap\\
    HDA & 45,075 & 0.516& 2.408& 0.000& 0.000& 16.445& Datacentermap\\
    lnPD & 45,075& 3.738& 1.701& -2.866& 3.761& 11.198& Census Bureau\\
    lnMHI & 45,075& 10.823& 0.280& 9.747& 10.812& 12.110& Census Bureau\\
    lnGDP & 45,075& 13.959& 1.609& 8.598& 13.777& 20.517& BEA\\
    YEAR & 15\\
    FIPS & 3,005\\
\bottomrule
\end{tabular}
\end{table*}
\subsection{Models}
This section outlines the empirical framework used to identify the causal effect of data centers on power capacity. To address high-dimensional confounding, the analysis proceeds in three steps. First, a Double Machine Learning (DML) framework is employed to provide debiased estimates of the average treatment effect, estimated across four alternative baseline specifications. Second, a dynamic event study framework, augmented with historical agglomeration (HDA) controls, is used to validate parallel trends and capture the precise timing of the grid response. Third, a placebo test with 500 random iterations is conducted to rule out spurious shocks and formally validate the causal identification strategy.

\textit{5.2.1 Double Machine Learning}. To estimate the causal effect of data center expansion on regional power capacity while flexibly accounting for high-dimensional confounding factors, this study employs the Double Machine Learning framework (Chernozhukov et al., 2018; Ahrens et al., 2025). Traditional linear regression suffers from regularization bias when dealing with a large number of controls. The DML approach overcomes this through orthogonalization (residualization) and cross-fitting. Estimating the nuisance functions and the treatment effect on the same sample induces overfitting bias. Five-fold cross-fitting partitions the data, estimates the nuisance functions on training folds, and computes residuals on held-out folds, ensuring approximate orthogonality and yielding debiased, reliable causal estimates. The baseline model is specified as follows: 
\begin{equation}
CAP_{i,t} = \theta_0 LDA_{j,t} + g(X_{j,t}) + \varepsilon_{j,t}
\end{equation}
where subscript i denotes the county and subscript t denotes the year. Capacity is the newly added power generation capacity; LDA represents the log of data center building area; X is a vector of high-dimensional covariates (including time-varying socioeconomic variables, county fixed effects, and year fixed effects); and g(X) is an unknown function estimated by a machine learning algorithm specifically, LassoCV. The covariate space is high-dimensional, including numerous county and year fixed effects together with socioeconomic controls. LassoCV automatically selects the relevant features and shrinks irrelevant coefficients to zero, thereby mitigating regularization bias that would otherwise arise in high-dimensional settings. The parameter of interest $\theta_0$, is estimated via a two-stage procedure. First, the nuisance functions E[CAP | X] and E[LDA | X] are estimated using K-fold cross-fitting. Second, the residuals from these regressions are regressed on each other to yield a debiased estimate of $\theta_0$. This study runs four primary specifications: (1) the standard main model, (2) a winsorized model to mitigate extreme value bias, (3) a lagged treatment model to establish temporal precedence, and (4) an alternative state-level fixed effects model. 
This study implements the Double Machine Learning estimator with five-fold cross-fitting. Specifically, the sample is randomly partitioned into five folds using KFold with shuffle=True and a fixed random seed of 42. In each iteration, two separate LassoCV models (with 3-fold internal cross-validation) are trained on the four training folds to residualize the outcome and the treatment; the residuals are then computed on the held-out test fold. This process is repeated so that every observation is used exactly once for residualization. Final inference is obtained via OLS on the orthogonalized residuals with heteroskedasticity-robust (HC1) standard errors. All estimations are conducted in Python using scikit-learn for the LassoCV nuisance models and statsmodels for the final residual-on-residual regression.

\textit{5.2.2 Event Study Design}. To examine the temporal dynamics of the treatment effect and validate the parallel trends assumption, this study extends the analysis to an event study framework (El Ghoul et al., 2022). The relative time to treatment is defined based on the first year a county’s data center area exceeds zero, with t = -1 set as the baseline period. To disentangle the causal effects of newly established data centers from persistent historical path dependence, an interaction term between historical data center agglomeration (HDA) and a linear time trend is incorporated into the nuisance feature space. The event study equation is specified as:
\begin{equation}
CAP_{i,t} = \sum \beta_k D_{j,t}^k + f(Z_{j,t}) + \mu_{j,t}
\end{equation}
where Dk is a dummy variable indicating the relative time period k (ranging from t = -5 to t = +5), and f(Z) represents the non-parametric nuisance function estimated by the machine learning algorithm, which absorbs HDA trends, socioeconomic controls, and two-way fixed effects. The coefficients of interest, $\beta_k$, measure the dynamic causal effect at each relative time period relative to the baseline period (t = -1).

\textit{5.2.3 Placebo Test}. To validate the robustness of the causal identification against spurious random shocks, a placebo test is conducted by randomly reassigning the timing of data center establishment across treated counties. The exact same DML procedure is iterated 500 times to generate a distribution of pseudo treatment effects under the null hypothesis of no true causal effect. If the true estimated treatment effect falls far outside the distribution of the randomized placebo estimates, the validity of the causal identification strategy is strongly confirmed (Eggers et al., 2024).

\section{ACKNOWLEDGMENTS}

The author gratefully acknowledges the U.S. Energy Information Administration for providing the generator level power capacity data, and the commercial digital infrastructure databases that supplied the data center building footprint information used in this study. The author also thanks the Development Planning Unit at University College London for institutional support.

\section{REFERENCES}

\begingroup
\renewcommand\bibsection{}
\nocite{*}
\bibliographystyle{ACM-Reference-Format}
\bibliography{sample-base}

@article{ahrens2025model,
  author  = {Ahrens, Achim and Hansen, Christian B. and Schaffer, Mark E. and Wiemann, Thomas},
  title   = {Model averaging and double machine learning},
  journal = {Journal of Applied Econometrics},
  year    = {2025},
  volume  = {40},
  number  = {3},
  pages   = {249--269},
  doi     = {10.1002/jae.3103}
}

@techreport{alvarez2026data,
  author      = {Alvarez, Fernando E. and Argente, David and Chow, Joyce and Van Patten, Diana},
  title       = {Data centers and local economies in the age of AI: A shift-share approach},
  institution = {National Bureau of Economic Research},
  year        = {2026},
  type        = {NBER Working Paper},
  number      = {35194},
  doi         = {10.3386/w35194}
}

@article{ancona2026urban,
  author  = {Ancona, Camilla and Lauber Bonomo, Ofek and Rozhkov, Anton and Porfiri, Maurizio},
  title   = {Urban infrastructure and fossil fuel industrial legacy drive US data center siting},
  journal = {Nature Cities},
  year    = {2026},
  volume  = {3},
  pages   = {756--767},
  doi     = {10.1038/s44284-026-00487-z}
}

@misc{aaf2025ai,
  author       = {{American Action Forum}},
  title        = {AI data centers: Which factors determine their location?},
  year         = {2025},
  howpublished = {\url{https://www.americanactionforum.org/insight/ai-data-centers-which-factors-determine-their-location/}},
  note         = {Accessed: 2026-10-06}
}

@article{ao2026data,
  author  = {Ao, Mingming and Cao, Yi and Loch, Christoph H. and Wan, Chi and Xi, Yaoyi},
  title   = {When data centers arrive: Data center development and local housing markets},
  journal = {SSRN},
  year    = {2026},
  pages   = {1--47},
  doi     = {10.2139/ssrn.6984701}
}

@article{arzumanyan2025geospatial,
  author  = {Arzumanyan, Mariam and Rodriguez Calzado, Edna and Lin, Ning and Bahadur, Vaibhav and Das, Jani and Ko, Tingwei Lucy and Koesterke, Lars},
  title   = {Geospatial suitability analysis for data center placement: A case study in Texas, USA},
  journal = {Sustainable Cities and Society},
  year    = {2025},
  volume  = {131},
  pages   = {106687},
  doi     = {10.1016/j.scs.2025.106687}
}

@misc{hanlin2026data,
  author       = {Hanlin, Michelle Isenhouer and Roehr, Andy},
  title        = {Data center impact on local communities: Balancing growth, infrastructure and community outcomes},
  year         = {2026},
  howpublished = {Bakertilly},
  url          = {https://www.bakertilly.com/insights/data-center-impact-on-local-communities},
  note         = {Accessed: 2026-10-06}
}

@incollection{baudry2021data,
  author    = {Baudry, Ken},
  title     = {Data center site search and selection},
  booktitle = {Data Center Handbook},
  edition   = {2nd},
  publisher = {Wiley},
  year      = {2021},
  chapter   = {22},
  doi       = {10.1002/9781119597537.ch22}
}

@misc{bluecap2026data,
  author       = {{Blue Cap Economic Advisors}},
  title        = {Data center site selection criteria every developer should evaluate},
  year         = {2026},
  howpublished = {BlueCap},
  url          = {https://www.bluecapeconomicadvisorsyu.com/post/data-center-site-selection-criteria and-every-developer-should Zhu-evaluate},
  note         = {Accessed: ,2026-10-06}
}

@misc{brookings2026turning,
  author       = {{Brookings Institution}},
  title        = {Turning the data center boom into long-term, local prosperity},
  year         = {2026},
  url          = {https://www.brookings.edu/articles/turning-the-data-center-boom-into-long-term-local-prosperity/},
  note         = {Accessed: 2026-10-06}
}

@misc{damicis2026how,
  author       = {Damicis, Jim and Tranmer, Alexandra},
  title        = {How do developers choose locations for new data centers?},
  year         = {2026},
  howpublished = {Camoin Associates},
  url          = {https://camoinassociates.com/resources/siting-and-building-data-centers/},
  note         = {Accessed: 2026-10-06}
}

@article{cao2024data,
  author  = {Cao, Zhiwei and Zhou, Xin and Wu, Xianghaomeng and Liu, Tracy and Neng, Jeffery and Wen, Yonggang},
  title   = {Data center sustainability: Revisits and outlooks},
  journal = {IEEE Transactions on Sustainable Computing},
  year    = {2024},
  volume  = {9},
  number  = {3},
  pages   = {236--248},
  doi     = {10.1109/TSUSC.2023.3281583}
}

@misc{cooper2026key,
  author       = {Cooper, Rob},
  title        = {Key factors to consider for effective data centre site selection},
  year         = {2026},
  howpublished = {CBRE UK},
  url          = {https://www.cbre.co.uk/insights/articles/key-factors-to-consider-for-effective-data-centre-site-selection},
  note         = {Accessed: 2026-10-06}
}

@article{chernozhukov2018double,
  author  = {Chernozhukov, Victor and Chetverikov, Denis and Demirer, Mert and Duflo, Esther and Hansen, Christian and Newey, Whitney and Robins, James},
  title   = {Double/debiased machine learning for treatment and structural parameters},
  journal = {The Econometrics Journal},
  year    = {2018},
  volume  = {21},
  number  = {1},
  pages   = {C1--C68},
  doi     = {10.1111/ectj.12097}
}

@misc{coolnet2025data,
  author       = {{Coolnet}},
  title        = {11 data center site selection criteria you should follow},
  year         = {2025},
  url          = {https://www.coolnetpower.com/news/11-data-center-site-selection-criteria-you-should-follow/},
  note         = {Accessed: 2026-10-06}
}

@misc{datax2024cs,
  author       = {{DataX Connect}},
  title        = {The 8Cs: What goes into choosing data centre locations?},
  year         = {2024},
  url          = {https://dataxconnect.com/insights-choosing-data-centre-locations/},
  note         = {Accessed: 2026-10-06}
}

@article{eggers2024placebo,
  author  = {Eggers, Andrew C. and Tu{\~n}{\'o}n, Guadalupe and Dafoe, Allan},
  title   = {Placebo tests for causal inference},
  journal = {American Journal of Political Science},
  year    = {2024},
  volume  = {68},
  number  = {3},
  pages   = {1106--1121},
  doi     = {10.1111/ajps.12818}
}

@article{elghoul2022event,
  author  = {El Ghoul, Sadok and Guedhami, Omrane and Mansi, Sattar A. and Sy, Oumar},
  title   = {Event studies in international finance research},
  journal = {Journal of International Business Studies},
  year    = {2022},
  volume  = {54},
  number  = {2},
  pages   = {344--364},
  doi     = {10.1057/s41267-022-00534-6}
}

@article{fang2025where,
  author  = {Fang, Tommy Pan and Greenstein, Shane},
  title   = {Where the cloud rests: The economic geography of data centers},
  journal = {Strategy Science},
  year    = {2025},
  volume  = {10},
  number  = {4},
  pages   = {404--420},
  doi     = {10.1287/stsc.2024.0225}
}

@article{fard2023platforms,
  author  = {Fard, Ali},
  title   = {Platforms and palimpsests: Urban landscapes of data in Northern Virginia},
  journal = {Footprint},
  year    = {2023},
  volume  = {17},
  number  = {2},
  pages   = {27--42},
  doi     = {10.59490/footprint.17.2.6726}
}

@misc{schaap2025how,
  author       = {Schaap, Andrew},
  title        = {How data centers are transforming local communities},
  year         = {2025},
  howpublished = {Fast Company},
  url          = {https://www.fastcompany.com/91301630/how-data-centers-are-transforming-local-communities},
  note         = {Accessed: 2026-10-06}
}

@misc{eusterman2026data,
  author       = {Eusterman, Michael},
  title        = {Data center site selection strategy guide},
  year         = {2026},
  howpublished = {Giga Energy},
  url          = {https://www.gigaenergy.com/blog/data-center-site-selection},
  note         = {Accessed: 2026-10-06}
}

@misc{goffri2026data,
  author       = {Goffri, Sholmy},
  title        = {Data center siting: Six factors that determine success},
  year         = {2026},
  howpublished = {Data Center Dynamics},
  url          = {https://www.datacenterdynamics.com/en/opinions/data-center-siting-six-factors-that-determine-success/},
  note         = {Accessed: 2026-10-06}
}

@misc{hardisty2026data,
  author       = {{Hardisty Jones}},
  title        = {Data centres and local socio-economic impacts: Jobs, infrastructure, and improving the attractiveness of places},
  year         = {2026},
  url          = {https://www.hardistyjones.com/insight/data-centres-and-local-socio-economic-impacts-jobs-infrastructure-and-improving-the-attractiveness-of-places/},
  note         = {Accessed: 2026-10-06}
}

@misc{harvard2026why,
  author       = {{Harvard Gazette}},
  title        = {Why are communities pushing back against data centers?},
  year         = {2026},
  url          = {https://news.harvard.edu/gazette/story/2026/04/why-are-communities-pushing-back-against-data-centers/},
  note         = {Accessed: 2026-10-06}
}

@article{huang2025site,
  author  = {Huang, Junya and Zhao, Ming and Yan, Jin and Zheng, Zhi},
  title   = {Site selection and design of big data centers},
  journal = {Telecommunications Science},
  year    = {2025},
  volume  = {41},
  number  = {2},
  doi     = {10.11959/j.issn.1000-0801.2025033}
}

@article{ma2026environmental,
  author  = {Ma, Yutong and Du, Jiaman},
  title   = {Environmental, social and economic justice of data centers on urban sustainability},
  journal = {npj urban sustainability},
  year    = {2026},
  doi     = {10.21203/rs.3.rs-10365556/v1}
}

@article{mongird2025more,
  author  = {Mongird, Kendall and Bracken, Cameron and Burleyson, Casey D. and Oikonomou, Konstantinos and Ou, Yang and Rice, Jennie S. and Thurber, Travis and Voisin, Nathalie},
  title   = {More land is needed for solar and wind infrastructure under a high renewables scenario in the Western US by 2050},
  journal = {Communications Earth \& Environment},
  year    = {2025},
  volume  = {6},
  pages   = {632},
  doi     = {10.1038/s43247-025-02632-3}
}

@article{xiao2025environmental,
  author  = {Xiao, Tianqi and Fuso Nerini, Francesco and Matthews, H. Damon and Tavoni, Massimo and You, Fengqi},
  title   = {Environmental impact and net-zero pathways for sustainable artificial intelligence servers in the USA},
  journal = {Nature Sustainability},
  year    = {2025},
  volume  = {8},
  pages   = {1541--1553},
  doi     = {10.1038/s41893-025-01681-y}
}

@article{rosati2023data,
  author  = {Rosati, Clayton and James, Aju and Metcalf, Kathryne},
  title   = {Data plantation: Northern Virginia and the territorialization of digital civilization in ``the Internet Capital of the World''},
  journal = {Online Media and Global Communication},
  year    = {2023},
  volume  = {2},
  number  = {2},
  pages   = {199--227},
  doi     = {10.1515/omgc-2023-0017}
}

@article{roychoudhury2026modeling,
  author  = {Roychoudhury, Saurav and Ross, Justin M.},
  title   = {Modeling economic impacts of a hyperscale data center campus},
  journal = {SSRN},
  year    = {2026},
  doi     = {10.2139/ssrn.7087540}
}

@article{siddik2021environmental,
  author  = {Siddik, Md Abu Bakar and Shehabi, Arman and Marston, Landon},
  title   = {The environmental footprint of data centers in the United States},
  journal = {Environmental Research Letters},
  year    = {2021},
  volume  = {16},
  number  = {6},
  pages   = {064017},
  doi     = {10.1088/1748-9326/abfba1}
}

@techreport{unu2026environmental,
  author      = {{United Nations University Institute for Water, Environment and Health (UNU-INWEH)}},
  title       = {Environmental cost of artificial intelligence: Carbon, water and land footprints},
  year        = {2026},
  doi         = {10.53328/INR26RMA002}
}

@article{vanzalk2018spatial,
  author  = {van Zalk, John and Behrens, P.},
  title   = {The spatial extent of renewable and non-renewable power generation: A review and meta-analysis of power densities and their application in the U.S.},
  journal = {Energy Policy},
  year    = {2018},
  volume  = {123},
  pages   = {83--91},
  doi     = {10.1016/j.enpol.2018.08.023}
}

@misc{deng2026what,
  author       = {Deng, Honghao},
  title        = {What will it take to make data centres more sustainable and fit for an AI future?},
  year         = {2026},
  howpublished = {World Economic Forum},
  url          = {https://www.weforum.org/stories/artificial-intelligence/build-data-centres-future-sustainable/},
  note         = {Accessed: 2026-10-06}
}

@misc{wri2026ways,
  author       = {{World Resources Institute (WRI)}},
  title        = {7 ways data centers affect US communities},
  year         = {2026},
  url          = {https://www.wri.org/insights/us-data-center-growth-impacts},
  note         = {Accessed: 2026-10-06}
}

@article{yue2026local,
  author  = {Yue, Daniel and Zeng, Yiyang},
  title   = {The local economic effects of data center entry},
  year    = {2026},
  doi     = {10.2139/ssrn.6497238}
}
\endgroup

\end{document}